\documentclass[prl,aps,twocolumn,preprintnumbers,showpacs,superscriptaddress]{revtex4}
\usepackage{amsfonts,amssymb,amsmath}            
\usepackage{graphics,graphicx,epsfig}            
\usepackage{amsthm}                              

\def\openone{\leavevmode\hbox{\small1\normalsize\kern-.33em1}}

\def\opone{\leavevmode\hbox{\small1\kern-3.8pt\normalsize1}}

\newcommand{\tr}[1]{\mbox{Tr} \, #1 }
\newcommand{\be}{\begin{equation}}
\newcommand{\ee}{\end{equation}}
\newcommand{\bea}{\begin{eqnarray}}
\newcommand{\eea}{\end{eqnarray}}

\begin{document}


\title{Direct measurement of non-linear properties of bipartite quantum states}


\author{Fabio Antonio \surname{Bovino}}%

\author{Giuseppe  \surname{Castagnoli}}%
\affiliation{Elsag spa, Via Puccini 2, 16154 Genova, Italy}

\author{Artur \surname{Ekert}}%
\affiliation{Centre for Quantum Computation, DAMTP,
             University of Cambridge,
             Cambridge CB3 0WA, UK}
\affiliation{Department of Physics,
             National University of Singapore,
             Singapore 117\,542, Singapore}

\author{Pawe{\l} \surname{Horodecki}}%
\affiliation{Faculty of Applied Physics and Mathematics,
            Gda\'nsk University of Technology,
            Gda\'nsk, Poland}

\author{Carolina \surname{Moura Alves}}%
\affiliation{Centre for Quantum Computation, DAMTP,
             University of Cambridge,
             Cambridge CB3 0WA, UK}

\author{Alexander Vladimir \surname{Sergienko}}%
\affiliation{Departments of ECE and Physics, Boston University,
             Boston, Massachusetts, USA}

\begin{abstract}
Non-linear properties of quantum states, such as entropy or
entanglement, quantify important physical resources and are
frequently used in quantum information science. They are usually
calculated from a full description of a quantum state, even though
they depend only on a small number parameters that specify the
state. Here we extract a non-local and a non-linear quantity, namely
the Renyi entropy, from local measurements on two pairs of
polarization entangled photons. We also introduce a ``phase marking"
technique which allows to select uncorrupted outcomes even with
non-deterministic sources of entangled photons. We use our
experimental data to demonstrate the violation of entropic
inequalities. They are examples of a non-linear entanglement
witnesses and their power exceeds all linear tests for quantum
entanglement based on all possible Bell-CHSH inequalities.
\end{abstract}

\maketitle


Many interesting properties of composite quantum systems, such as
entanglement or entropy, are not measured directly but are
inferred, usually from a full specification of a quantum state
represented by a density operator. However, it is interesting to
note that some of these properties can be measured in the same way
we measure and estimate average values of observables. Here we
illustrate this by extracting a non-local quantity, the Renyi
entropy of the composite system, from local measurements on two
pairs of polarization entangled photons. This quantity is a
non-linear function of the density operator. We then use our
experimental data to demonstrate the violation of entropic
inequalities, which can be also interpreted as the experimental
demonstration of a non-linear entanglement witness.

Consider a source which generates pairs of photons. The photons in
each pair fly apart from each other to two distant locations A and
B. Let us assume that the polarization of each pair is described
by some density operator $\varrho$, which is unknown to us.
Following Schr\"odinger's remarks on relations between the
information content of the total system and its
sub-systems~\cite{Schr35}, it has been proven that for separable
states global von Neumann entropy is always not less then local
ones~\cite{RPH94}. Subsequently a number of entropic inequalities
have been derived, satisfied by all separable
states~\cite{HHH96,CA97,Ter02,VW02}. The simplest one is based on
the Renyi entropy, or the purity measure, $\tr(\varrho^{2})$ and
can be rewritten as
\begin{equation}
\tr(\varrho_{A}^{2})\geq \tr(\varrho^{2}),
\tr(\varrho_{B}^{2})\geq \tr(\varrho^{2}), \label{nonlin}
\end{equation}
where $\varrho_{A}$ and $\varrho_{B}$ are the reduced density
operators pertaining to individual photons. The
inequalities~(\ref{nonlin}) involve non-linear functions of
density operators and are known to be stronger than \emph{all}
Bell-CHSH inequalities~\cite{CHSH,HHH96}. There are entangled
states which are not detected by the Bell-CHSH inequalities but
which are detected by the entropic inequalities~(\ref{nonlin}).
\footnote{For example, if we mix the maximally entangled singlet
state with the maximally depolarized state, in proportions $p$ and
$1-p$ respectively ($0\le p \le 1$), then the resulting state is
entangled for $p>\frac{1}{3}$. However, the entanglement is
detected by the CHSH inequalities for $p>\frac{1}{\sqrt 2}\approx
0.707107$ and by the non-linear inequalities for $p>\frac{1}{\sqrt
3}\approx 0.57735$.}

\begin{figure}
\includegraphics[angle=0, width=.50\textwidth]{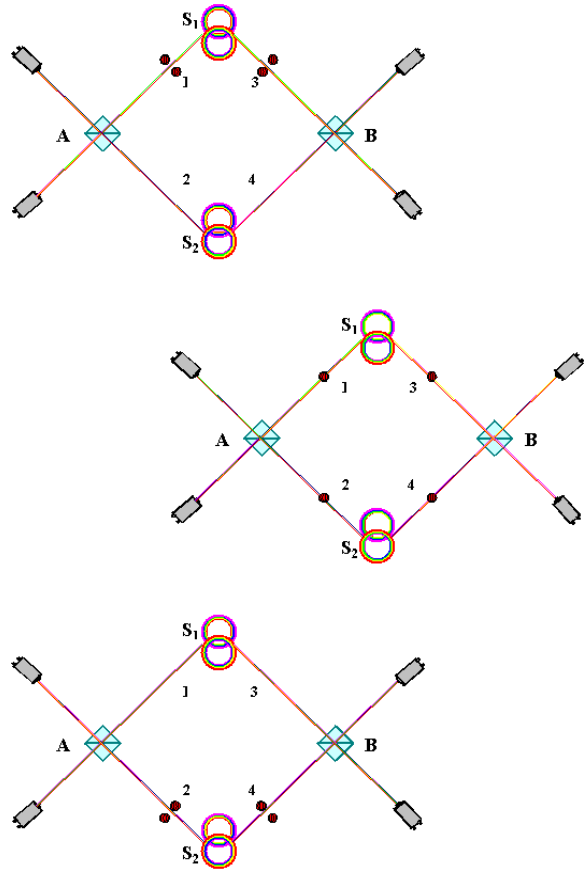}
\caption{Possible emissions leading to four-photons coincidences.
Parametric down-conversion is not an ideal source of entangled
photons. It generates a superposition of vacuum, two-entangled
photons, four-entangled photons, etc. The central diagram shows the
desired emission of two independent entangled pairs -- one by source
$S_1$ and one by source $S_2$. The top and the bottom diagrams show
unwelcome emissions of four photons by one of the two sources.
Although the three cases are equally likely we can eliminate
contributions of unwelcome emissions by a judicious choice of the
phase difference between the beams that coherently pump the two BBO
crystals. We call it ``phase marking". Symmetric superposition of
the unwelcome emissions, obtained for $\phi =0$, does not contribute
to the coalescence-anticoalescence events, whereas the antisymmetric
superposition, obtained for $\phi =\pi /2$ does not affect the
anticoalescence-anticoalescence outcomes.}\label{possiblepaths}
\end{figure}

In the experiment a source generates pairs of
polarization-entangled photons in a singlet state $\left\vert
H\right\rangle \left\vert V\right\rangle -\left\vert
V\right\rangle \left\vert H\right\rangle $, where $H$ and $V$
stand for horizontal and vertical polarizations respectively.
\begin{figure}
\includegraphics[angle=0,width=.45\textwidth]{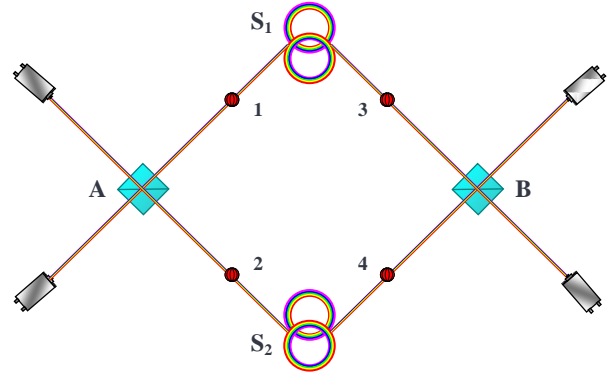}
\caption{An outline of our experimental set-up. Sources $S_1$ and
$S_2$ emit pairs of polarization-entangled photons. The entangled
pairs are emitted into spatial modes $1$ and $3$, and $2$ and $4$.
One photon from each pair is directed into location A and the
other into location B. At the two locations photons impinge on
beam-splitters and are then detected by photodetectors. There are
four possible outcomes in this experiment: coalescence at A and
coalescence at B, coalescence at A and anti-coalescence at B,
anti-coalescence at A and coalescence at B, anti-coalescence at A
and anti-coalescence at B. The probabilities associated with the
four outcomes are, respectively, $p_{cc}$, $p_{ca}$, $p_{ac}$,
$p_{aa}$ (subscript \emph{b} stands for coalescence and \emph{a}
for anticoalescence). In terms of these probabilities the entropic
inequalities~(\ref{nonlin}) are written as $p_{ca}\geq p_{aa}$,
$p_{ac}\geq p_{aa}$.}\label{outline}
\end{figure}
\begin{figure*}
\includegraphics[angle=-90,width=.50\textwidth]{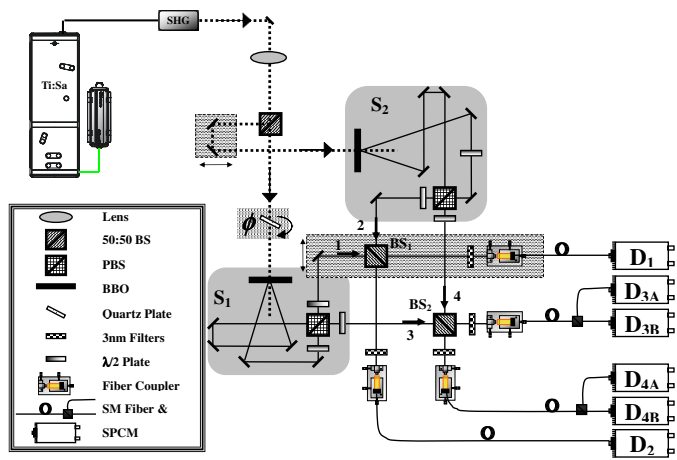}
\caption{Experimental set-up. Two pairs of polarization entangled
photons are generated in two separate 3mm thick BBO crystals via
typeII spontaneous parametric down conversion (SPDC). The crystals
are pumped by two beams, each consisting of a train of femtosecond
laser pulses with the central wavelength of 415nm and a repetition
rate of 76MHz (12.5ns between pulses). The beams are displaced by
few micrometers using a trombone and the sub wavelength path
difference (phase shift $\phi $) is introduced by a tilted 1mm
thick quartz-plate inserted into one of the pump beams. The two
pump beams are obtained by splitting the main beam (average power
470mW), which is generated by doubling the frequency of the output
of a Coherent MIRA-VERDI laser (operating in the mode-locking
regime, 830nm central wavelength, 9nm bandwidth, and 160fs of
pulse duration) using second harmonic generation (SHG) in a
2mm-thick LBO crystal. Entangled photons from the BBO crystals
enter interferometers which provide high fidelity polarization
entangled states in ultra-fast regime~\cite{BCDC04}. Subsequently
the photons are sent to the beam-splitters $BS_1$ and $BS_2$, so
that each beam-splitter receives two photons, one photon from each
entangled pair. The photons are coupled in Single-Mode (SM)
Fibers~\cite{Bov03} and detected by the silicon-based
single-photon counting modules (Perkin-Elmer SPCM-AQR-14). Once
the correct position of the two beam splitters is established the
four-fold coincidence interference patterns are measured.
Coincidence counts $D_{1}\otimes D_{2}\otimes D_{3A}\otimes
D_{4A}$, $ D_{1}\otimes D_{2}\otimes D_{3A}\otimes D_{4B}$,
$D_{1}\otimes D_{2}\otimes D_{3B}\otimes D_{4A}$ and $D_{1}\otimes
D_{2}\otimes D_{3B}\otimes D_{4B}$ allow to estimate the rate of
anticoalescence-anticoalescence events, whereas $ D_{1}\otimes
D_{2}\otimes D_{3A}\otimes D_{3B}$ and $D_{1}\otimes D_{2}\otimes
D_{4A}\otimes D_{4B}$ give the anticoalescence-coalescence
rate.}\label{expdetails}
\end{figure*}
Thus $\rho $ is a maximally entangled pure state and,
consequently, $\rho _{A}=\rho _{B}$ are maximally mixed,
completely depolarized states. In order to measure a quadratic
property of $\rho $ we need to operate on at least two copies of
the entangled pairs. Here we use the phenomenon of coalescence and
anti-coalescence of photons. If two identical photons are incident
on two different input ports of a beam-splitter they will coalesce
i.e. they will emerge together in one of the two, randomly chosen,
output ports. More precisely, all pairs of photons (in general all
pairs of bosons) with a symmetric (under the exchange of photons)
polarization state will coalesce and all pairs of photons with an
antisymmetric polarization state will anti-coalesce i.e. photons
will emerge separately in two different output ports of the
beam-splitter. A beautiful experimental observation of this effect
was reported by Hong, Ou, and Mandel over fifteen years
ago~\cite{Mandel}, and more recently by~Di Giuseppe et
al~\cite{sashabunching}. The main idea behind our experiment is
illustrated in Fig.~(\ref{outline}).Two independent pairs of
photons are generated by sources $S_1$ and $S_2$, and one photon
from each pair is directed into location A and the other into
location B. At the two locations the photons impinge on
beam-splitters and are then detected by photo-detectors. The
beam-splitters at A and B, as long as the photons from two
different pairs arrive within the coherence time, effectively
project on the symmetric and anti-symmetric subspace in the four
dimensional Hilbert space associated with the polarization degrees
of freedom. Let us consider the four possible detections in this
experiment: coalescence at A and coalescence at B, coalescence at
A and anti-coalescence at B, anti-coalescence at A and coalescence
at B, and finally, anti-coalescence at A and anti-coalescence at
B. Let the probabilities associated with the four outcomes be,
respectively $p_{cc}$, $p_{ca}$, $p_{ac}$, $p_{aa}$, and
(subscript \emph{c} stands for coalescence and \emph{a} for
anti-coalescence). In technical terms they correspond to
probabilities of projecting the state $\rho \otimes \rho$ of two
pairs of photons on symmetric or antisymmetric subspaces at
location A (associated with spatial modes $1$ and $2$) and B
(associated with spatial modes $3$ and $4$), e.g. $p_{ca}=Tr\left(
P_{S}\otimes P_{A}\right) \left( \rho \otimes \rho \right)$ etc,
where $P_{S}$ and $P_{A}$ are the corresponding projectors on the
symmetric and antisymmetric subspaces. We can now write
\begin{eqnarray}
\tr\varrho^2 &=& p_{cc}- p_{ca}-p_{ac}+p_{aa},\\
\tr\varrho^2_A &=& p_{cc}+ p_{ca}-p_{ac}-p_{aa},\\
\tr\varrho^2_B &=& p_{cc}- p_{ca}+p_{ac}-p_{aa},
\end{eqnarray}
and the inequalities (\ref{nonlin}) can be expressed in a new and
simple form,
\begin{equation}
p_{ca}\geq p_{aa}\quad, \quad p_{ac}\geq p_{aa} \label{prob}
\end{equation}
Theoretical predictions for the singlet state are $p_{cc}=3/4$, $
p_{ac}=p_{ca}=0$ and $p_{aa}=1/4$. Let us mention in passing that
in this particular case a coincidence count at A (B) projects the
state of the remaining two photons at B (A) on the singlet state,
inducing an entanglement swapping, c.f.~\cite{PBWZ98}. The
polarization entangled photons are generated using type-II
spontaneous parametric down conversion (SPDC)~\cite{Kwiat}. An
ultraviolet pulse from a pump laser is split into two pulses which
are slightly delayed with respect to each other and directed
towards two $\beta $-barium-borate (BBO) crystals. The two BBO
crystals correspond to the two sources $S_1$ and $S_2$. When the
pulses pass through the crystals they emit, with some probability,
pairs of energy-degenerate polarization-entangled photons into
modes 1 and 3 (source $S_1$), and 2 and 4 (source $S_2$). Modes 1
and 2 are coupled by the beam-splitter at A and modes 3 and 4 by
the beam-splitter at B. Behind the beam-splitters silicon
single-photon counting modules register emerging photons. The
coalescence and anti-coalescence coincidences are recorded.  (See
Fig.~(\ref{expdetails}) for technical details.) Currently
available sources of entangled photons are probabilistic which
creates an additional experimental difficulty. When a UV pulse
passes through a BBO crystal it produces a superposition of
vacuum, two entangled photons, four entangled photons, etc. A
four-photon coincidence in our set-up may be caused by two
entangled pairs from two different sources but also by four
photons from one source and no photons from the other, as shown in
Fig.~(\ref{possiblepaths}), moreover, the three scenarios are
equally likely. In order to discriminate unwelcome four-photon
coincidences we have used ``phase marking" - for certain values of
the phase difference between the two pumping beams we register
only coincidences that not corrupted by the spurious emissions.
The description can be made more quantitative by analysing an
effective Hamiltonian describing entanglement generation in two
coherently pumped BBO crystals,
\begin{equation}
H=\eta (K+K^{\dagger })+\eta (Le^{-i\phi }+L^{\dagger }e^{i\phi }).
\end{equation}
Here $\eta $ is a coupling constant, proportional to the amplitude
of the pumping beams, $\phi $ is the relative phase shift between
the beams introduced by the tilted quartz-plate, and
$K=a_{1H}a_{3V}-a_{1V}a_{3H}$ and $L=a_{2H}a_{4V}-a_{2V}a_{4H}$ are
the linear combination of annihilation operators describing the
down-converted modes. The subscripts $1,2,3,4$ label the spatial
modes and $H$, $V$ stand for horizontal and vertical polarizations.
The four-photon term of a quantum state generated by this
Hamiltonian can be written as
\begin{eqnarray}
\left\vert \,\Psi \right\rangle  &=&\frac{e^{i\phi }}{\sqrt{10}}
(a_{1H}^{\dagger }a_{3V}^{\dagger }-a_{1V}^{\dagger
}a_{3H}^{\dagger })(a_{2H}^{\dagger }a_{4V}^{\dagger
}-a_{2V}^{\dagger }a_{4H}^{\dagger })
\notag \\
&+&\frac{1}{\sqrt{10}}\left( \frac{1}{2}a_{1H}^{\dagger
2}a_{3V}^{\dagger 2}-a_{1H}^{\dagger }a_{1V}^{\dagger
}a_{3V}^{\dagger }a_{3H}^{\dagger }+
\frac{1}{2}a_{1V}^{\dagger 2}a_{3H}^{\dagger 2}\right)   \notag \\
&+&\frac{e^{2i\phi }}{\sqrt{10}}\left( \frac{1}{2}a_{2H}^{\dagger
2}a_{4V}^{\dagger 2}-a_{2H}^{\dagger }a_{2V}^{\dagger }a_{4V}^{\dagger
}a_{4H}^{\dagger }+\frac{1}{2}a_{2V}^{\dagger 2}a_{4H}^{\dagger 2}\right)
\notag \\
&&\left\vert \,vac\right\rangle ,\label{totalstate}
\end{eqnarray}
where the first term describes the desired two
polarization-entangled pairs, each in the singlet state
$\left\vert \,H\right\rangle \left\vert \,V\right\rangle
-\left\vert \,V\right\rangle \left\vert \,H\right\rangle $,
whereas the last two terms describe unwelcome four-photon states
generated by an emission from only one of the two crystals (see
Fig.\ref{possiblepaths}).
The coalescence and anti-coalescence coincidences for the
state~(\ref{totalstate}) are given by
\begin{equation}
p_{ac}=p_{ca}=\frac{3}{20}(1-\cos 2\phi ),p_{aa}=\frac{1}{4}+\frac{3}{20}%
\cos 2\phi .
\end{equation}
In order to recover the coincidences associated with the desired
singlet state we notice that for $\phi =0$ and $\phi =\pi /2$
there are no spurious contributions to $p_{ac}=p_{ca}$ and
$p_{aa}$ respectively. For these two phase settings the symmetric
and antisymmetric superposition of the last two terms
in~(\ref{totalstate}) lead to additional symmetries at the input
of the beam-splitters and cancels out the unwelcome outcomes.

\begin{figure}
\includegraphics[width=.48\textwidth]{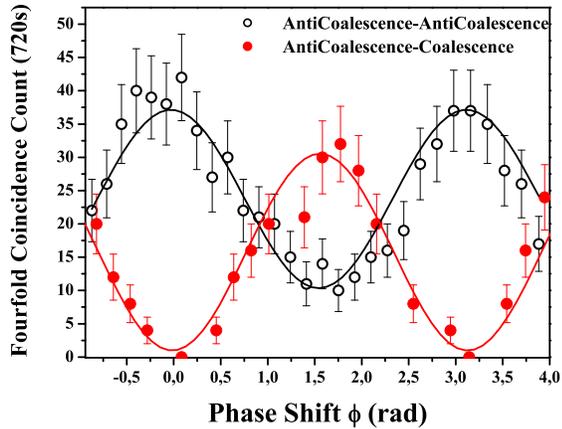}
\caption{Four-photon coincidences.  The plot shows the four-photon
interferences as a function of the pump displacement which is
proportional to the relative phase shift $\phi$ of the two pumps. The probability of coalescence and anti-coalescence for ideal singlet state case can be extracted from the coincidence counts at the minima of interference curves using (8).
The minima of the curves correspond to the required values of
$p_{ac}$ and $p_{aa}$. The experimental data show that
$p_{aa}>p_{ac}$, which violates the entropic
inequalities~(\ref{nonlin}). The results for the curve
anticoalescence-coalescence are normalized by a factor 2 to take
into account the fact that half of the coincidences are
lost.}\label{data}
\end{figure}

In the experiment we have traced the dependence of $p_{ac}(\phi)$
and $p_{aa}(\phi)$. Even though the measurement of $p_{ac}=p_{ca}$
and $p_{aa}$ could only be realized for different values of the
phase $\phi$, the difference in probability values is clearly
observed by following the minima of both curves in
Fig.~(\ref{data}). Based on the statistical fit of curves in Fig.~(\ref{data}) we obtain $p_{ac}=p_{ca}= 0.0255 \pm 0.008$ and $p_{aa} = 0.2585 \pm 0.008$. The measured result is in good agreement with the
theoretical prediction and clearly demonstrates the violation of the
entropic inequalities for the singlet state.
Let us stress that these inequalities involve nonlinear functions
of a quantum state. Their power exceeds all linear tests such as
the Bell-CHSH inequalities with all possible settings and
entanglement witnesses. In fact our result can be viewed as the
first experimental measurement of a non-linear entanglement
witness. Direct measurements of non-linear properties of quantum
states open new ways of probing and manipulating quantum
phenomena.
\begin{acknowledgments}
The experiment has been carried out in the Quantum Optics Laboratory
of ELSAG spa, Genova, Italy, within the project "Quantum
Cryptographic Key Distribution" co-funded by the Italian Ministry of
Education, University and Research (MIUR), Grant No. 67679/L.488 and EC-FET project QAP-2005-015848.
This work has been also supported by the Cambridge-MIT Institute,
the Fujitsu Laboratories Europe and the EC (Project
RESQ No. IST-2001-37559). A.E. acknowledges financial support of the
A*Star Grant No. 012-104-0040. P.H. acknowledges support from the
Polish Ministry of Scientific Research, Information and Technology
(Project No. PBZ-Min-008/P03/03). C.M.A. is supported by the
Funda{\c c}{\~a}o para a Ci{\^e}ncia e Tecnologia (Portugal). AVS
acknowledges support by the National Science Foundation (NSF) and by
the Defence Advanced Research Project Agency (DARPA). F.~A.~B. acknowledges help of Sig. G. Tarantino and machine shop at ELSAG spa.
\end{acknowledgments}

\end{document}